\documentclass[showpacs,showkeys,amsmath,amssymb,aps,onecolumn,pre]{revtex4}
\usepackage{graphicx}
\usepackage{pifont}
\usepackage{dcolumn}
\usepackage{bm}

\begin{document}
\preprint{}
\title[Properties of nano-islands formation]{Properties of nano-islands formation in nonequilibrium reaction-diffusion systems with memory effects}
\author{Vasyl O.~Kharchenko}
\email{vasiliy@ipfcentr.sumy.ua}
\affiliation{Institute of Physics, University of Augsburg, Universit\"{a}t Str. 1, D-86135 Augsburg, Germany}
\affiliation{Institute of Applied Physics, National Academy of Sciences of Ukraine, 58 Petropavlovskaya St., 40030 Sumy, Ukraine} 

\author{Dmitrii O.~Kharchenko, Sergei V.~Kokhan, Irina V.~Vernigora}
\affiliation{Institute of Applied Physics, National Academy of Sciences of Ukraine, 58 Petropavlivska St., 40030 Sumy, Ukraine}

\author{Vladimir V.Yanovsky}
\affiliation{Institute for Single Crystals, National Academy of Sciences of Ukraine, 60 Lenin ave., 61178 Kharkiv, Ukraine}

\begin{abstract}
We study dynamics of pattern formation in systems belonging to class of
reaction-Cattaneo models including persistent diffusion (memory effects of the
diffusion flux). It was shown that due to the memory effects pattern seletion
process are realized. We have found that oscillatory behavior of the radius of
the adsorbate islands is governed by finite propagation speed. It is shown that
stabilization of nano-patterns in such models is possible only by
nonequilibrium chemical reactions. Oscillatory dynamics of pattern formation is
studied in details by numerical simulations. \\
\end{abstract}
\maketitle

\section{Introduction}

From theoretical and experimental observations it is known that
reaction-diffusion systems play an important role in the study of generic
spatiotemporal behavior of nonequilibrium systems. Usually such models admit
main contributions related to both local dynamics (chemical reactions type of
birth-and-death processes) and mass transport. Novel experimental methods, such
as field ion microscopy, scanning tunneling microscopy can be used to monitor
chemical reactions on the metal surfaces with atomic resolution.

In adsorption-desorption processes when material can be deposited from the
gaseous phase such experimental methods allow one to investigate formation of
clusters or islands of adsorbed molecules/atoms \cite{ZTWE96}. Such islands can
have linear size of nanometer range \cite{GLRBE94}. In
Refs.\cite{KNSZGC91,PWCL97,BGBK98,CF99,Nature99} it was experimentally shown
that nano-patterns on solid surface and nano-islands in adsorbed mono-atomic
layers can be organized. It was found that patterns on scales shorter than the
diffusion length emerge from the interplay of reactions and lateral
interactions between adsorbed particles. The adsorbate presence can modify the
local crystallographic structures of the substrate's surface layer producing
long range interactions between adsorbed atoms and their clusters (see for
example, Refs.\cite{ZKRGL94,M91,V92}). It was observed experimentally that
nanometer-sized vacancy islands can be organized in a perfect triangular
lattice when a single monolayer of Ag was exposed on Ru(0001) surface at room
temperature \cite{Nature99}. Nanometer elongated islands was observed
experimentally in Si/Si(100) \cite{202}, the same was found at deposition of Ge
on Si \cite{sem2002}, metallic elongated islands were observed at deposition of
Cu on Pd(110) \cite{206}. It was shown that elongated adsorbate clusters are
governed by formation of dimers and their reconstructions \cite{MBE1998}
representing nonequilibrium chemical reactions.

It is well known that short-range transient patterns can be observed at initial
stages of phase separation processes \cite{Binder}. These transient patterns
can be stabilized by nonequilibrium chemical reactions leading to emergence of
stationary patterns \cite{TH96}. In models of systems with adsorption and
desorption processes additional chemical reactions should be introduced to
freeze the patterns. It was shown previously, that adsorption and thermal
desorption are ``equilibrium reactions'' which can not induce the formation of
kinetic spatially modulated stationary phases \cite{HM96,BHKM97}. A problem of
formation of stationary microstructures in such systems with irreversible
nonequilibrium chemical reactions was considered in
Refs.\cite{HME98_1,HME98_2}. Properties of pattern formation in systems of
adsorption-desorption type with dissipative dynamics were studied previously
\cite{ME94,BHKM97}. Pattern formation in pure dissipative stochastic systems
with multiplicative noise obeying fluctuation-dis\-sipation relation was
discussed in Refs.\cite{MW2005,M2010,PhysD2009,PhysScr2011}. It was shown that
such multiplicative noise can sustain stationary patterns of nano-size range in
pure dissipative systems.

As far as the reactions are defined through chemical kinetics, an evolution of
the field variable, let say coverage for adsorption/desorption systems, is
governed by the reaction-diffusion equation. A typical deterministic equation
is of the form
\begin{equation}\label{eqX}
\partial_t x= f(x)-\nabla\cdot \mathbf{J},
\end{equation}
where $x=x(\mathbf{r},t)$ is the local coverage at surface defined as the
quotient between the number of adsorbed particles in a cell of the surface and
the fixed number of available sites in each cell, $x\le 1$. The term $f(x)$
stands for local dynamics and describes birth-and-death or
adsorption-desorption processes; the flux $\mathbf{J}$ represents the mass
transport.

Most of the theoretical studies deal with the standard Fick law
$\mathbf{J}=-D\nabla x$, where $D$ is the diffusion constant. It is known that
at $f=0$ the ordinary diffusion equation $\partial_tx=\nabla\cdot D\nabla x$
has the unrealistic feature of infinitely fast (infinite) propagation. In such
a case a fundamental solution $$x(r,t)=(4\pi Dt)^{-1/2}\exp(-r^2/Dt)$$ means
that for any small $t$ at any large $r$ the quantity $x$ will be nonzero,
though exponentially small. It leads to unphysical effect that particles move
with infinite speed (more than sound speed in solids). The reason for this is
lack of correlations of particle motion. To avoid such pathology the diffusion
flux can be generalized by taking into account memory effects \cite{JP89}
\begin{equation}\label{eqJ0}
 \mathbf{J}=-\int\limits_0^tM(t,t')D\nabla x(\mathbf{r},t'){\rm d}t'
\end{equation}
described by the memory kernel $M(t,t')=\tau_J^{-1}\exp(-|t-t'|/\tau_J)$. The
delay time $\tau_J$ is related to correlated (persistent) random walk. In the
limit $f(x)=0$ one gets the finite propagation speed $\sqrt{D/\tau_J}$. At
$\tau_J\to 0$ the asymptotic $M(t,t')=\delta (t-t')$ leads to the Fick law
$\mathbf{J}=-D\nabla x$ with infinite propagation. As far as real systems
(molecules, atoms) have finite propagation speed one should use Eq.(\ref{eqJ0})
or an equivalent equation: $\tau_J\partial _t\mathbf{J}=-\mathbf{J}-D\nabla x$
\cite{H1999}. Equations (\ref{eqX},\ref{eqJ0}) can be combined into the one
reaction-Cattaneo equation of the form
\begin{equation}\label{hypX}
\tau_J\partial^2_{tt} x+(1-\tau_J f'(x))\partial_{t}x= f(x)+\nabla\cdot D\nabla
x,
\end{equation}
where prime denotes derivative with respect to the argument. In the absence of
reaction term $(f=0)$ the corresponding telegraph equation has a solution of
the form
\begin{equation}\nonumber
x(r,t)=\left\{\begin{split}
 &\frac{1}{N}\exp\left(-\frac{t}{2\tau_J}\right)I_0\left(\sqrt{\frac{\chi}{N}}\right),\
 \
\text{for}~~ |r|<\sqrt{\frac{D t}{\tau_J}},\\
 &0 \ \ \  \text{otherwise},
\end{split}\right.
 \end{equation}
where $N=\sqrt{4 D\tau_J}$, $\chi=(D/\tau_J)t^2-r^2$, $I_0(\cdot)$ is the
modified Bessel function. As was pointed out in Ref.\cite{H1999} this equation
has some restrictions: (i) it typically does not preserve positivity of the
solution $x(\mathbf{r},t)$; (ii) the damping coefficient must be positive,
i.e., $f'(x)<\tau_J^{-1}$.

Therefore, one can use more realistic model given by Eq.(\ref{hypX}), where
particles have finite speed at smaller time scales and approach diffusion
motion on larger time scales. It follows that depending on the form of the
reaction term $f(x)$ Eq.(\ref{hypX}) admits oscillatory solutions \cite{H1999}.
An application of such formalism for phase separation processes study with
finite $\tau_J$ allows one to describe pattern selection processes at early
stages of decomposition in binary systems (see Refs.\cite{lzg,PhysA1_2010}) and
oscillatory formation of the ordered phase in crystalline systems \cite{CEJP}.
Oscillatory solutions in class of reaction-Cattaneo systems with fluctuating
quantity $\tau_J$ were considered in Ref.\cite{Ghosh2010}.

In this study we are aimed to describe dynamics of pattern formation and
selection processes in a class of reaction-Cattaneo systems given by
Eq.(\ref{hypX}). Novelty of our approach is in studying oscillatory dynamics of
pattern formation in such class of models with chemical reactions governed by
adsorption/desorption processes. Following the formalism proposed in
Refs.\cite{HME98_2,MW2005,M2010} we compare a behaviour of ordinary
reaction-diffusion dissipative system and reaction-Cattenao system. It will be
shown below that for the last class of models pattern selection processes are
realized, stable patterns possible only if nonequilibrium chemical reactions
are introduced. Studying behavior of islands size as clusters of dense phase we
will show that averaged island size behaves itself in oscillatory manner.
Considering formation of islands of adsorbed particles we discuss properties of
the island size distribution during the system evolution.

The paper is organized as follows. In Section II we propose the stochastic
reaction-Cattenao model. The linear stability analysis and properties of
pattern selection processes are presented in Section III. We discuss results of
numerical simulation in Section IV. Finally, in Section V, we draw conclusions
from our study.

\section{Model}

Let us consider a model where only one class of particles is possible.
Following Refs.\cite{MW2005,M2010,ME94,BHKM97,CWM2002,HME98_1,HME98_2} one
assumes that the particles can be adsrobed, desorbed, can diffuse and interact
among themselves. Therefore, we introduce the scalar field describing dynamics
of the local coverage at surface $x(\mathbf{r},t)\in[0,1]$. The reaction term
incorporating adsorption and desorption terms together with nonequilibrium
chemical reactions is as follows: $f_0(x)=k_a
p(1-x)-k_dx\exp(U(\mathbf{r})/T)-k_rx^n$. Here $k_a$ and $k_d$ are adsorption
and desorption rates, respectively; $p$ is the partial pressure of the gaseous
phase; $U(\mathbf{r})$ is the interaction potential. The last term corresponds
to nonequilibrium chemical reaction of the order $n\ge1$ with the rate constant
$k_r$. Usually it estimates islands/dimers formation or associative desorption
\cite{MBE1998}. In further consideration we put $n=2$.

The total flux $\mathbf{J}$ is a sum of both ordinary diffusion flux
$(-D_0\nabla x)$ and flow of adsorbate $(-(D_0/T) x(1-x)\nabla U)$. Here the
multiplier $x(1-x)$ denotes that the flux is only possible to the $(1-x)$ free
sites. Hence, the total flux is
\begin{equation}
\mathbf{J}=-D_0\nabla x-\frac{D_0}{T}x(1-x)\nabla U;
\end{equation}
in an equivalent form one has
\begin{equation}\label{J1}
\mathbf{J}=-D_0 M(x)\left[\frac{\nabla x}{x(1-x)}+\frac{1}{T}\nabla U\right],
\end{equation}
where the Cahn mobility $M(x)=x(1-x)$ is introduced.

Formally the right hand side of Eq.(\ref{J1}) can be rewritten as follows:
\begin{equation}
\mathbf{J}=-D_0 M(x)\nabla \frac{\delta \mathcal{F}}{\delta x},
\end{equation}
where the total mesoscopic free energy functional is
\begin{equation}
\mathcal{F}=\mathcal{F}_0+\mathcal{F}_{int}.
\end{equation}
The non-interacting part takes the form
\begin{equation}
 \mathcal{F}_0=\int{\rm d}\mathbf{r}
 \left[x(\mathbf{r})\ln(x(\mathbf{r}))+(1-x(\mathbf{r}))\ln(1-x(\mathbf{r}))\right],
\end{equation}
whereas $\mathcal{F}_{int}$ is governed by the interaction potential $U$ which
we assume in the standard form \cite{BHKM97}
\begin{equation}
 U(\mathbf{r})=-\int{\rm d}\mathbf{r}' u(\mathbf{r}-\mathbf{r}')x(\mathbf{r}'),
\end{equation}
where $-u(r)$ is the binary attraction potential for two adsorbate particles
separated by the distance $r$, it is of symmetrical form , i.e. $\int{\rm
d}\mathbf{r}\mathbf{r}^{2n+1}u(\mathbf{r})=0$, $n=1,\ldots,\infty$.

Therefore, one can rewrite the total mesoscopic free energy as follows:
\begin{equation}\label{Ftot}
\mathcal{F}=\int{\rm d}\mathbf{r}\left[x\ln(x)+(1-x)\ln(1-x)\right]-
\frac{1}{2T}\iint{\rm d}\mathbf{r}{\rm d}\mathbf{r}'
x(\mathbf{r})u(\mathbf{r}-\mathbf{r}')x(\mathbf{r}').
\end{equation}

Following Ref.\cite{HME98_2} as a simple approximation for the interaction
potential, we choose the Gaussian profile
\begin{equation}\label{Gu}
u(r)=\frac{2\epsilon}{\sqrt{4\pi r_0^2}}\exp(-r^2/4r_0^2),
\end{equation}
where $\epsilon$ is the interaction strength, $r_0$ is the interaction radius.
Assuming that $x$ does not vary significantly within the interaction radius,
one can estimate
\begin{equation}\label{expansionU}
\int{\rm d}\mathbf{r}' u(\mathbf{r}-\mathbf{r}')x(\mathbf{r}')\simeq \int{\rm
d}\mathbf{r}'u(\mathbf{r}-\mathbf{r}')\sum_n
\frac{(\mathbf{r}-\mathbf{r}')^n}{n!}\nabla^n x(\mathbf{r}).
\end{equation}
Substituting Eq.(\ref{Gu}) into Eq.(\ref{expansionU}) up to terms of the 4-th
order one gets $\int u(r)x(r){\rm d}r=2\epsilon x$, $\frac{1}{2!}\int u(r)
r^2\nabla^2 x(r){\rm d}r=2\epsilon r_0^2\nabla^2x$, $\frac{1}{4!}\int u(r)
r^4\nabla^4 x(r){\rm d}r=\epsilon r_0^4\nabla^4 x$.

Therefore, using notation $\varepsilon=\epsilon/T$ one has
\begin{equation}
\frac{1}{T}\int{\rm d}\mathbf{r}'
u(\mathbf{r}-\mathbf{r}')x(\mathbf{r}')\simeq\varepsilon
x(\mathbf{r})+\varepsilon(1+r_0^2\nabla^2)^2x(\mathbf{r}).
\end{equation}
The total free energy functional takes the form\footnote{Here we deal with the
mesoscopic free energy functional $\mathcal{F}[x]$ not the true thermodynamic
free energy $F=-T\ln\int{\rm D}x\exp(-\mathcal{F}[x]/T)$. From the
methodological viewpoint one can note that the interaction part in
$\mathcal{F}$ have different sign as phase field crystals approach predicts
(see Refs.\cite{Grant200220061,Grant200220062,Grant200220063}).}
\begin{equation}
\label{Ftot1} \mathcal{F}=\int{\rm
d}\mathbf{r}\left[-\frac{\varepsilon}{2}x^2+x\ln
x+(1-x)\ln(1-x)-\frac{\varepsilon}{2}x(1+r_0^2\nabla^2)^2x \right].
\end{equation}
Therefore, the total flux can be written as
\begin{equation}
\mathbf{J}=-D_0M(x)\nabla\left[\frac{\delta\mathcal{F}_{loc}}{\delta
x}-\varepsilon(1+r_0^2\nabla^2)^2x\right],
\end{equation}
here $\mathcal{F}_{loc}$ is defined through the local part of the free energy
density (first three terms in Eq.(\ref{Ftot1})).

Next, it will be more convenient to measure time in units $k_d$, introduce the
diffusion length $L_{d}=\sqrt{D_0/k_d}$ and dimensionless adsorption rates
$\alpha=k_a p/k_d$, $\beta=k_r/k_d$. Therefore, the reaction term takes the
form $f(x)=\alpha(1-x)- x e^{-2\varepsilon x}-\beta x^2$ and the system is
described by two length scales\footnote{ For $r_0$ one has estimation $r_0\sim 1nm$, 
whereas $L_{dif}\sim 1 \mu m$ \cite{HME98_1,HME98_2}.}, where $r_0\ll L_{d}$.
As far as real systems (molecules, atoms) have finite propagation speed one
should take into account memory (correlation) effects, assuming
\begin{equation}\label{eqJ}
 \mathbf{J}=-L_d\int\limits_0^t{\rm d}t'\mathcal{M}(t,t';\tau_J) \nabla \frac{\delta\mathcal{F}}{\delta
 x(\mathbf{r},t')},
\end{equation}
where $\mathcal{M}(t,t')$ is the memory kernel. Taking it in the exponential
decaying form
$\mathcal{M}(t,t')=\tau_J^{-1}M(x(\mathbf{r},t'))\exp(-|t-t'|/\tau_J)$, where
$\tau_J$ is the flux relaxation time instead of one equation for the coverage
we get a system of two equations:
\begin{equation}
\begin{split}
\partial_t &x=f(x)-L_d\nabla\cdot\mathbf{J};\\
\tau\partial_t &\mathbf{J}=-\mathbf{J}-L_d
M(x)\nabla\frac{\delta\mathcal{F}}{\delta x},\quad \tau=\tau_J k_d.
\end{split}
\end{equation}
At $\tau_J\to 0$ the asymptotic $M(t,t')=\delta (t-t')$ leads to the Fick law
$\mathbf{J}=-L_d M(x)\nabla{\delta\mathcal{F}_{tot}}/{\delta x}$ with an
infinite propagation.

The equivalent equation for the coverage takes the form
\begin{equation}\label{Det}
\begin{split}
 &\tau\partial^2_{tt}x+\gamma(x)\partial_t x=\varphi(x,\nabla),\quad
\gamma(x)\equiv 1-\tau f'(x);\\ &\varphi(x;\nabla)\equiv
f(x)+L_d^2\nabla\cdot[\nabla x-\varepsilon M(x)(\nabla x+\nabla
\mathcal{L}_{SH}x)],\quad \mathcal{L}_{SH}=(1+r_0^2\nabla^2)^2.
\end{split}
\end{equation}

\begin{figure}[!t]
\centering
 a) \includegraphics[width=0.45\textwidth]{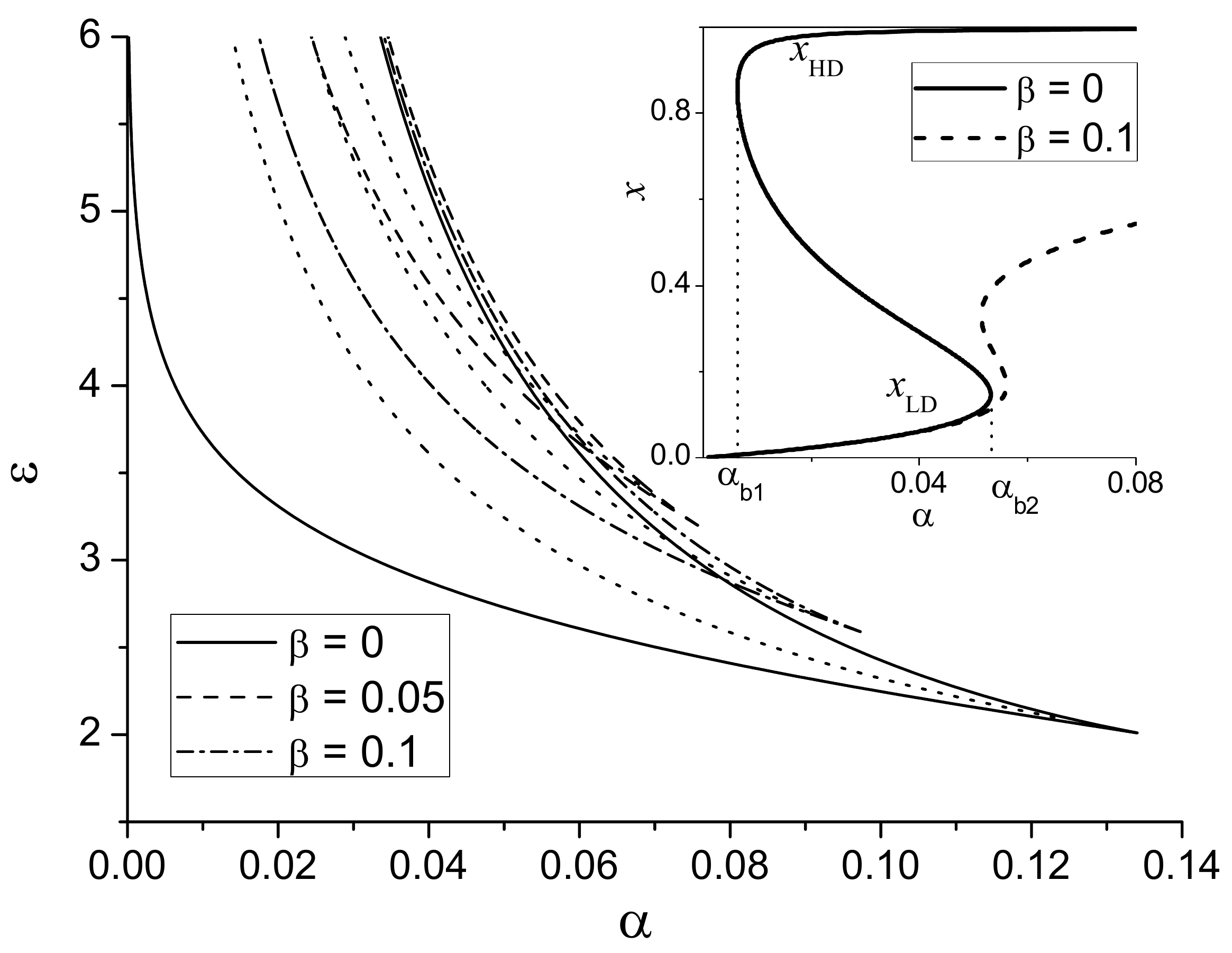}
 b) \includegraphics[width=0.45\textwidth]{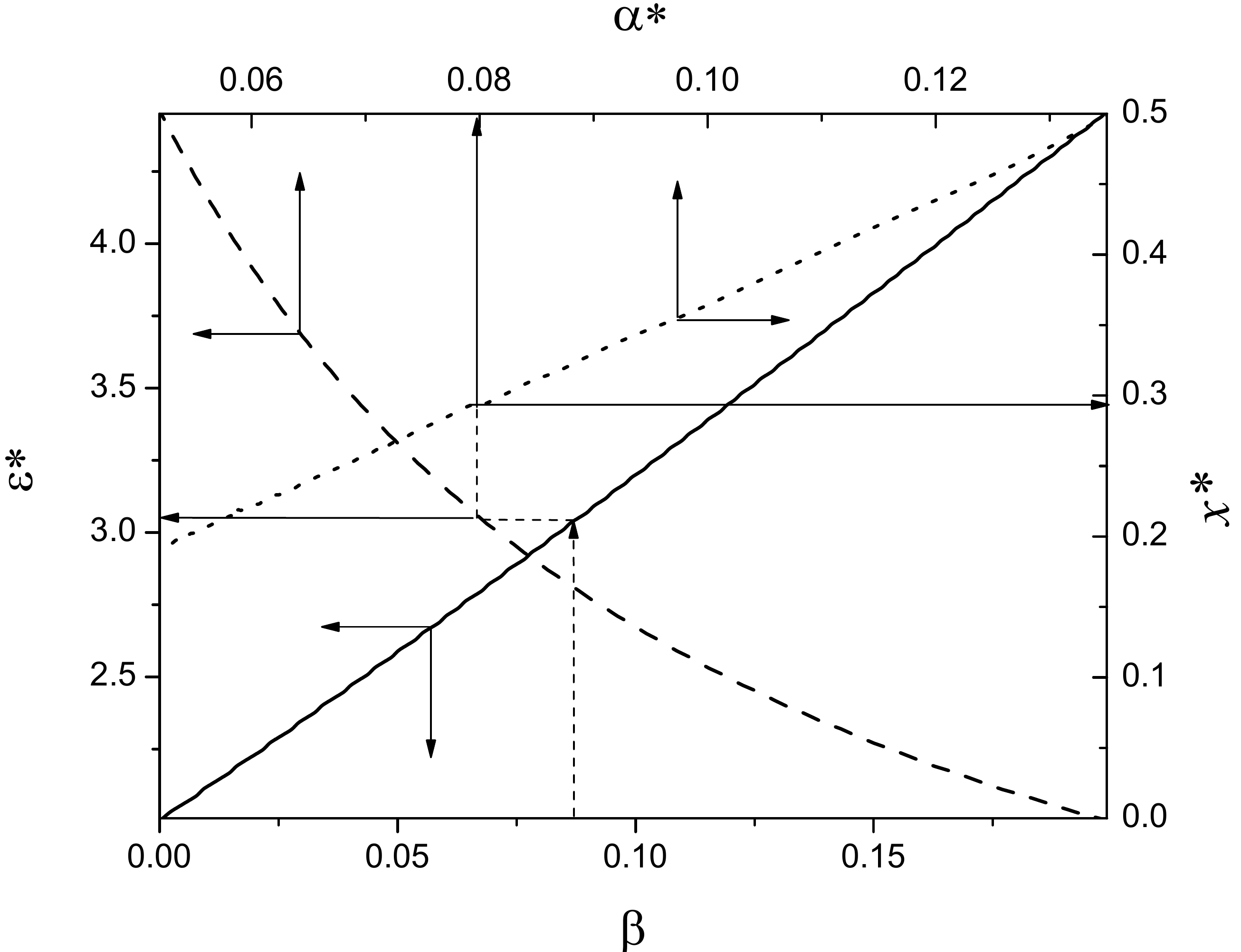}
 \caption{Phase diagram for homogeneous system in the parameter plane
$(\alpha,\varepsilon)$ (a). Dependencies $x(\alpha)$ in insertion are obtained
at $\varepsilon=4$. Critical points located at $\varepsilon^*$ and $\alpha^*$,
and the corresponding coverage $x^*$ are shown in plot(b)\label{uniform}}
\end{figure}

Let us consider stationary homogeneous system states. In the deterministic
limit in the absence of reaction, the system undergoes first order phase
transition, where its stationary uniform states are given by
$\alpha(1-x)=x\exp(-2\varepsilon x)$. The critical point of this equilibrium
phase transition $x_c=1/2$ is located at $\varepsilon_c=2$ and
$\alpha_c=\exp(-2)$. A coexistence line of diluted $(x<1/2)$ and dense
$(x>1/2)$ phases is given by the relation $\alpha=\exp(-\varepsilon)$.
Introduction of the chemical reactions governed by the rate $\beta$ shifts the
whole phase diagram and shrinks the domain where the system is bistable. Here
possible values for $x$ related to the uniform states decreases whereas
critical values for $\varepsilon$ become larger; if $\beta$ grows then critical
values for $\alpha$ decrease. The corresponding phase diagram is shown in
Fig.\ref{uniform}. Outside the cusp the system is monostable: at small $\alpha$
(before the cusp) the system is in low density state $x_{LD}$, whereas at large
$\alpha$ the high density state $x_{HD}$ is realized. In the cusp the system is
bistable. Critical points and the related coverage are shown in
Fig.\ref{uniform}b. Here starting from the fixed value for $\beta$ and moving
up to an intersection with the curve $\varepsilon^*(\beta)$, one gets the
critical value $\varepsilon^*$ as the corresponding ordinate in the left axis.
Than, we move to the right hand side to an intersection with the curve
$\varepsilon^*(\alpha^*)$; the corresponding abscissa in the top correspond to
critical value $\alpha^*$. Moving from this point to an intersection the curve
$x^*(\alpha^*)$, one gets the critical value $x^*$ as the corresponding
ordinate in the right axis. Acting in such a manner, we obtain all needed
critical values for the system parameters. Analytical relation between
$\alpha^*$, $\varepsilon^*$, $x^*$ and $\beta$ are as follows:
$\varepsilon^*=(1+\sqrt{5-2 x^*})/(2x^*(2-x^*))$,
$\alpha^*=2\varepsilon^{*2}x^{*3}\exp(-2\varepsilon^* x^*)$,
$\beta=2\varepsilon^*(1-\varepsilon^*x^*)\exp(-2\varepsilon^*x^*)$.

\section{Linear stability analysis}

It is known that systems with memory effects admit pattern selection processes
at fixed set of the system parameters \cite{lzg,PhysA1_2010,CEJP}. These
processes can be observed at early stages of the system evolution where linear
effects are essential. Therefore, pattern selection can be studied considering
stability of statistical moments, reduced to the averaged filed and/or
structure function as the Fourier transform of a two-point correlation function
for the coverage. As far as fourth order contribution in the interaction
potential $u(r)$ is not essential at small $r_0\ll L_d$, next we consider a
case where $\int u(r)x(r){\rm d}r\simeq 2\varepsilon(1+r_0^2\nabla^2)x$,
neglecting $r_0^4\nabla^4 x$.

Averaging Eq.(\ref{Det}) over initial conditions and taking $\langle
x\rangle-x_0\propto e^{{\rm i}(\omega t-kr)}$ one gets the dispersion relation
of the form
\begin{equation}\label{dr}
 \omega(k)_\mp=-\frac{{\rm i}\gamma(x_0)}{2\tau}\mp
 \left[
	 \frac{L_d^2k^2(1-2\varepsilon
M(x_0)(1-r_0^2k^2))-f'(x_0)}{\tau} -\frac{\gamma^2(x_0)}{4\tau^2}\right]^{1/2}.
\end{equation}
One can see that $\omega(k)$ can have real and imaginary parts, i.e.
$\omega(k)=\Re \omega(k)\pm {\rm i}\Im\omega(k)$. The component $\Re \omega(k)$
is responsible for oscillatory solutions, whereas $\Im\omega(k)$ describes
stability of the solution $\langle \delta x_\mathbf{k}(\omega)\rangle$.
Analysis of both $\Re \omega(k)$ and $\Im\omega(k)$ allows us to set a
threshold for a wave-number where oscillatory solutions are possible. Moreover,
it gives a wave-number for the first unstable solution. From the obtained
dispersion relations it follows that at $k=k_0$ satisfying equation
\begin{equation}
k_0^2(1-2\varepsilon M(x_0)(1-r_0^2k_0^2))=\frac{1}{L_d^2}\left[f'(x_0)
+\frac{\gamma^2(x_0)}{4\tau}\right]
\end{equation}
two branches of the dispersion relation degenerate. The unstable mode appears
at $k=k_c$ obtained from the equation
\begin{equation}\label{kc}
L_d^2k_c^2(1-2\varepsilon M(x_0)(1-r_0^2k_c^2))=f'(x_0).
\end{equation}
From the dispersion relation $\omega(k)$ one can find the most unstable mode
$k_m$ as a solution of the equation ${\rm d}\Im\omega/{\rm d}k=0$. It coincides
with first unstable mode when only one nonzeroth solution of the equation
$\Im\omega(k)=0$ emerges.

\begin{figure}
\centering
\includegraphics[width=0.5\textwidth]{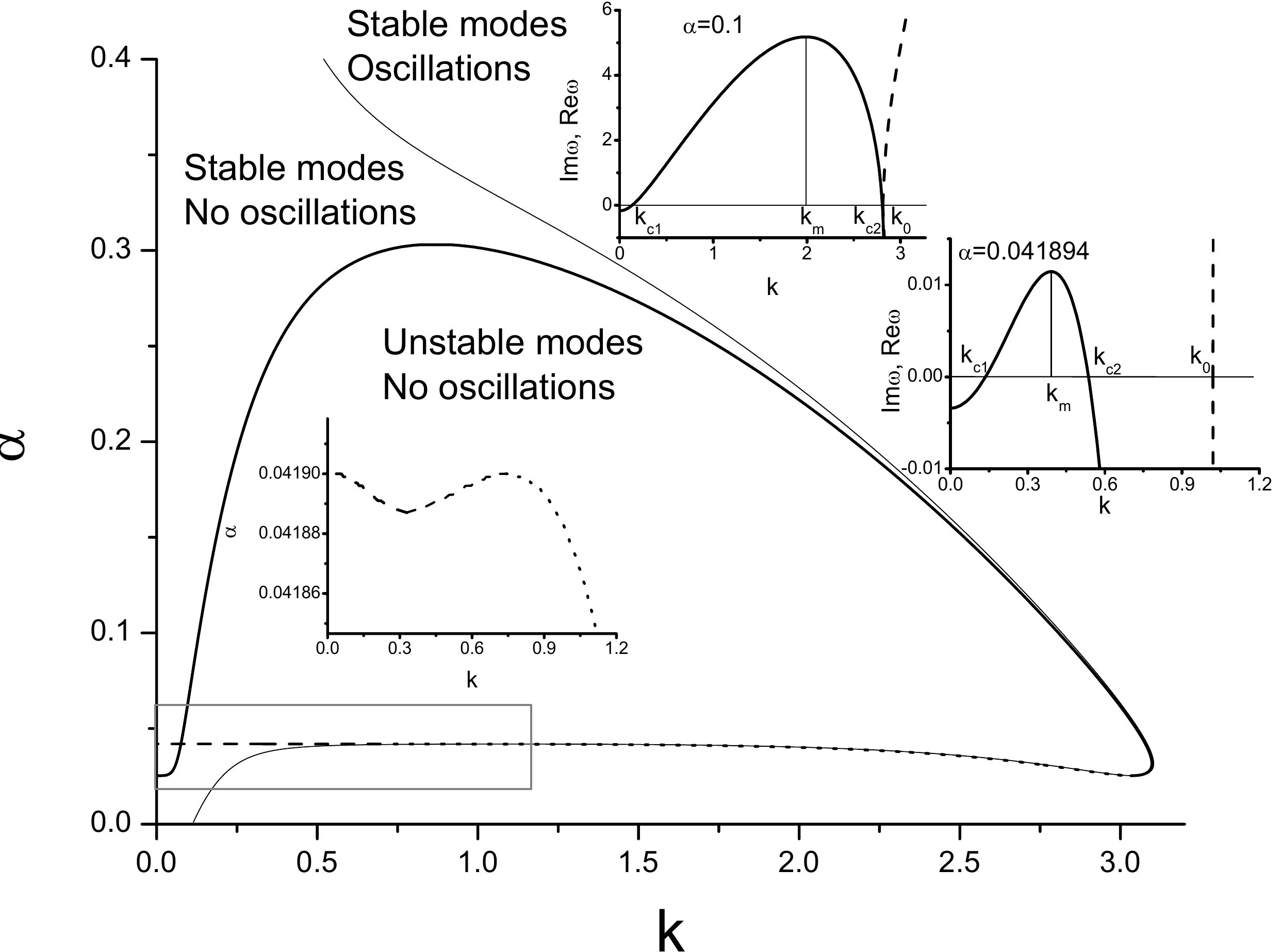}
\caption{Stability diagram at $\varepsilon=5$, $\tau=0.5$, and $\beta=0.1$.
Dependencies of $\Im \omega(k)$ and $\Re\omega(k)$ are shown as solid and
dashed lines in insertions at $\alpha=0.1$ and $\alpha=0.041894$ related to
high density and low density phases\label{a(k)}}
\end{figure}
Using obtained relations one can calculate a diagram indicating spatial
stability of all homogeneous states to inhomogeneouse perturbations. The
corresponding diagram is shown in Fig.\ref{a(k)}. Here domain of unstable modes
with respect to inhomogeneous perturbations is limited by solid and dashed
thick curves. The solid curve relates to high density phase, whereas dashed
line corresponds to low density phase; dotted line addresses to unstable
homogeneous stationary state. When we increase the adsorption rate $\alpha$
from zeroth value the first unstable mode emerges at large $k$ and is possible
only for high density phase. There is small domain for $\alpha$ where spatial
instability of the low density phase is possible (see magnified insertion of
$\alpha(k)$ at small $\alpha$ and $k$). It is should be noted that wave numbers
related to these unstable modes in both low- and high density phases are
observed in fixed interval $k\in [k_{c1},k_{c2}]$. The thin solid line in
Fig.\ref{a(k)} denotes critical values $k_o$ where oscillatory solutions
$\langle x(k,t)\rangle$ are possible. The domain of unstable modes with respect
to inhomogeneous perturbations is limited by large values for $\alpha$. It
means that instability of high density phase is possible only in fixed interval
for adsorption rate values.

\begin{figure}
\centering
 a) \includegraphics[width=0.45\textwidth]{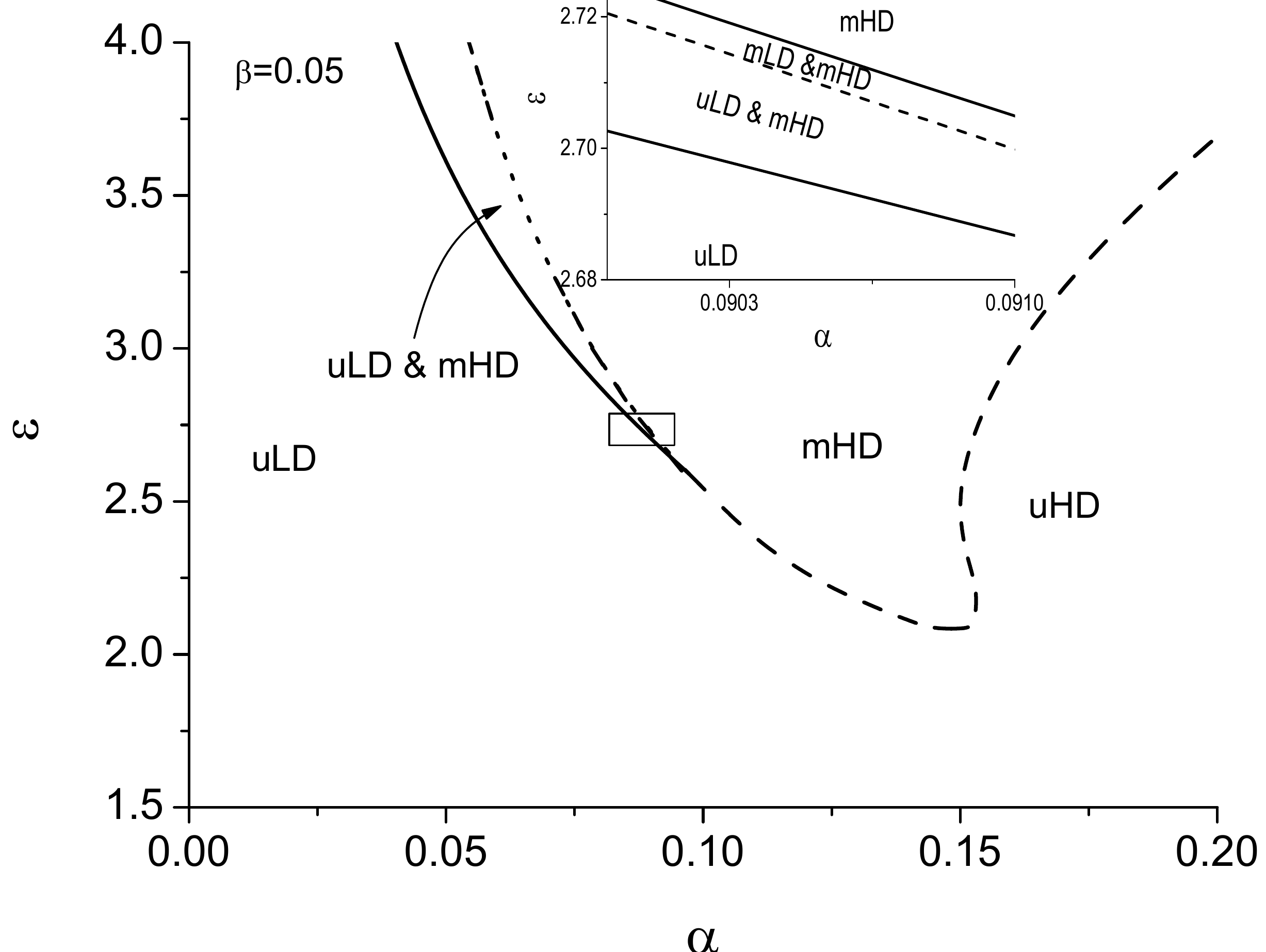}
 b) \includegraphics[width=0.45\textwidth]{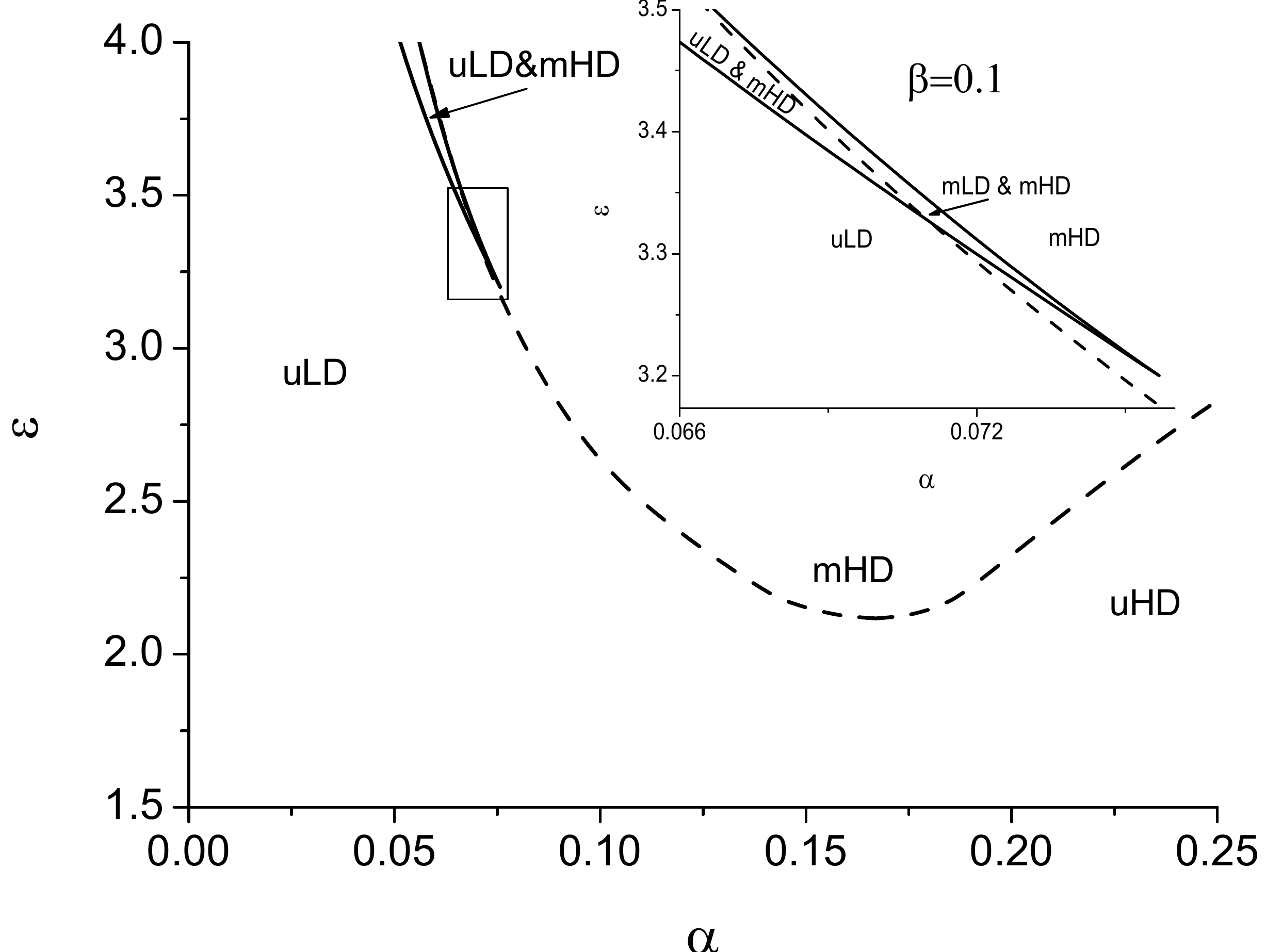}
\caption{Critical values for $\varepsilon$ and $\alpha$ related to pattern
formation: a) $\beta=0.05$; b) $\beta=0.1$ \label{ea_st}}
\end{figure}

According to obtained dependencies $\alpha(k)$ we plot in Fig.\ref{ea_st}
critical values for $\alpha$ and $\varepsilon$ related to formation of
spatially modulated phases, where solid lines denote binodals, dashed lines
bound spatially modulated phases.
In the case of $\beta=0$ one has only binodals bounding domains of one uniform
phase and domain of two uniform phase existence. Here no spatially modulated
phases are possible. When we put $\beta\ne0$ and increase the adsorption rate
at small $\alpha$ only uniform low density state (uLD) is possible; at large
$\alpha$ one has only uniform high density state (uHD). In the cusp depending
on the values for $\alpha$ one has: uniform low density state and modulated
high density phase (uLD\& mHD) at small $\alpha$; modulated low density and
high density phases (mLD\& mHD) at elevated $\alpha$. If we further increase
the adsorption rate only modulated high density phase (mHD) is possible.
Controlling the rate $\beta$ one can govern size of the domain where both
modulated low density and high density phases emerge; when $\beta$ grows the
domain of the modulated high density phase increases and mHD phase can emerge
before the first binodal (at smaller $\alpha=\alpha_{b1}$).

Considering properties of pattern selection we need to find dynamical equation
for the structure function $S(\mathbf{k},t)$ as the Fourier transform of the
two-point correlation function $\langle\delta x(\mathbf{r},t)\delta
x(\mathbf{r}',t)\rangle$ and study its behavior at small times. To that end we
obtain the linearized evolution equation for the Fourier components $\delta
x_{\mathbf{k}}(t)$ and $\delta x_{-\mathbf{k}}(t)$ and compute
$S(\mathbf{k},t)=\langle\delta x_\mathbf{k}(t)\delta
x_{-\mathbf{k}}(t)\rangle$. The corresponding dynamical equation takes the form
\begin{equation}
\tau\partial^2_{tt}S(\mathbf{k},t)+\gamma(x_0)\partial_tS(\mathbf{k},t)=2\left\{f'(x_0)-L_d^2k^2(1-2\varepsilon
M(x_0)(1-r_0^2k^2))\right\}
  S(\mathbf{k},t).
\end{equation}
Analytical solution can be found assuming $S(\mathbf{k},t)-S_0\propto
\exp(-i\varpi(\mathbf{k})t)$, where
\begin{equation}
\varpi(k)_\pm=-\frac{{\rm i}\gamma(x_0)}{2\tau}\pm\left[\frac{2\left(L_d^2
k^2(1-2\varepsilon
M(x_0)(1-r_0^2k^2))-f'(x_0)\right)}{\tau}-\frac{\gamma^2(x_0)}{4\tau^2}\right]^{1/2}.
\end{equation}
As in the previous case $\Im\varpi$ is responsible for stability of the system,
whereas $\Re\varpi$ relates to pattern selection processes.

\section{Numerical simulations}

A morphology of emergent patterns and dynamics of pattern formation we study by
numerical simulations in two dimensional system with $256\times256$ sites and
periodic boundary conditions. In our simulations we take time step $\Delta
t=2.5\times 10^{-4}$ and consider the case when $L_d=40 r_0$. The total size of
the system is $L=12.8L_d$. As initial conditions we take: $\langle
x(\mathbf{r},0)\rangle=0$, $\langle (\delta x(\mathbf{r},0))^2\rangle=0.1$
where $x(\mathbf{r},t)\in[0,1]$.

\begin{figure}[!t]
\centering
\includegraphics[width=120mm]{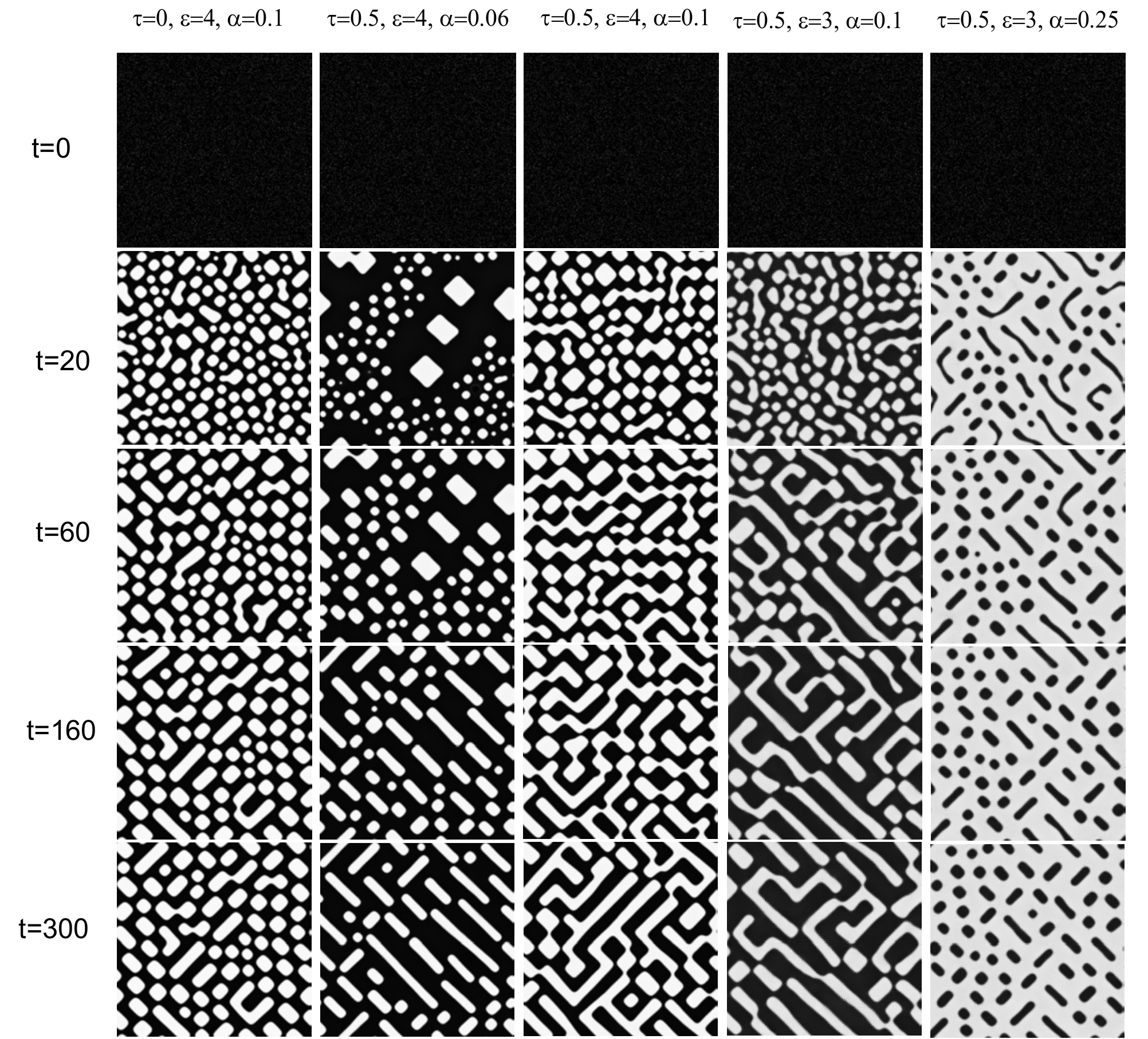}
\caption{Snapshots of the system evolution at $\beta=0.1$. All snapshots are
taken at $t=0$, $t=20$, $t=60$, $t=160$, $t=300$. Here concentration of adatoms
is shown with the help of gray scale: white domains correspond to adatoms
present, dark ones indicate regions without adatoms. \label{evol}}
\end{figure}

Typical evolution of the system with different values for $\tau$, $\varepsilon$
and $\alpha$ at $\beta=0.1$ is shown in Fig.\ref{evol}. In the case of pure
dissipative system ($\tau=0$) at $\alpha=0.1$ the adsorbate is organized into
separated islands with small difference in the linear sizes of islands. When we
put $\tau=0.5$ (see 3-rd column) at the same other system parameters we get
pattern where all islands at large time interval are combined into one
percolating cluster. Fixing $\tau=0.5$ and taking small $\alpha$ one gets set
of adsorbate islands with large difference in their linear sizes (see 2-nd
column). With further increase in $\alpha$ islands of vacancies are formed (see
4-th column). Such islands has the same structure as islands of adsorbate at
small adsorption rate.

\begin{figure}\centering
 \includegraphics[width=0.6\textwidth]{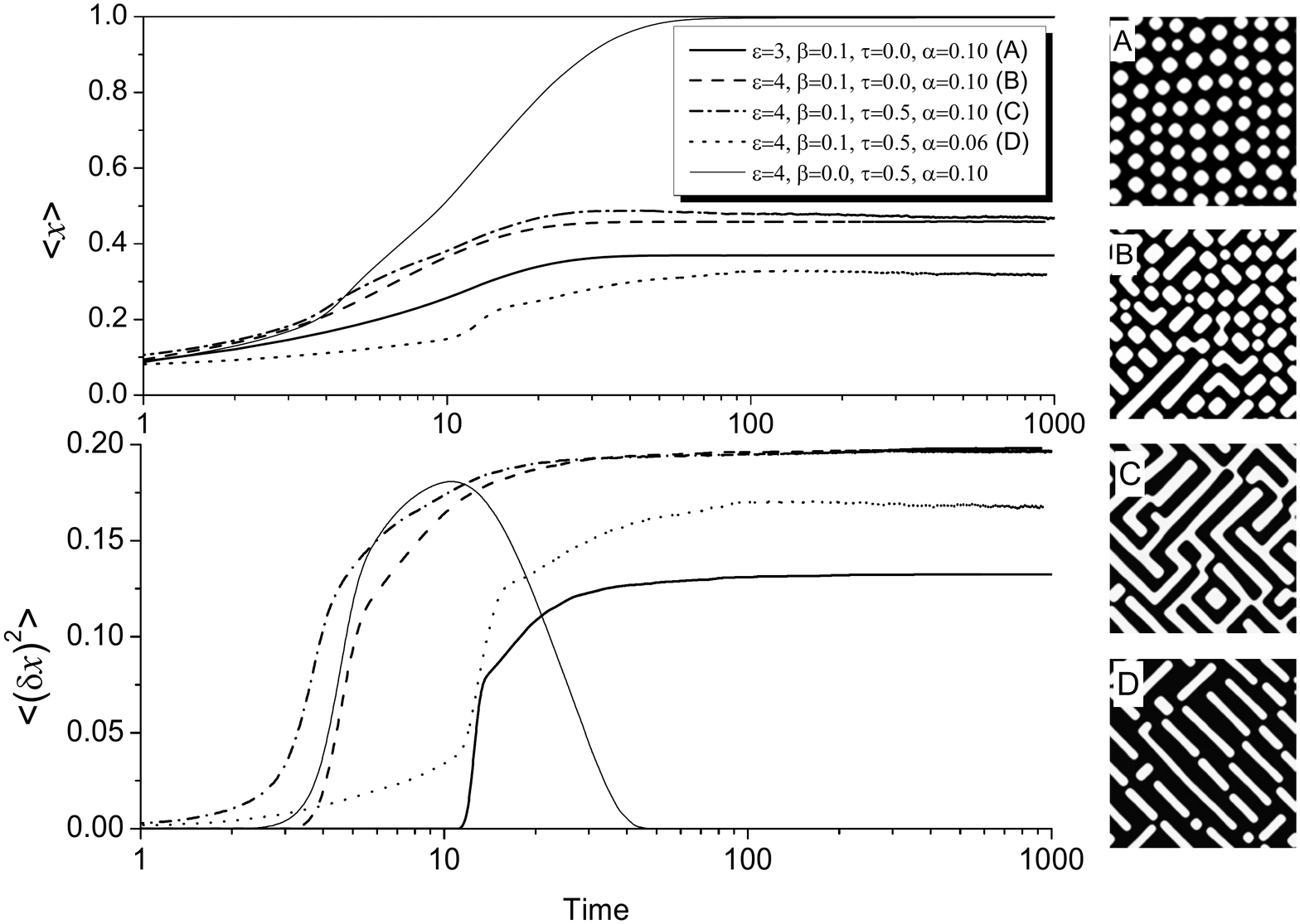}
 \caption{Evolution of $\langle x\rangle$ and $\langle(\delta x)^2\rangle$\label{avX}}
\end{figure}

In Figure \ref{avX} we present dynamics of the averaged concentration field
$\langle x\rangle$ and its dispersion $\langle(\delta x)^2\rangle$. It is known
that the growth of the quantity $\langle(\delta x)^2\rangle$ means ordering of
the system (an increase in fluctuations of the concentration). Following the
obtained dependencies one can find that if nonequilibrium reactions are absent
($\beta=0$) the system passes toward an equilibrium thermodynamic state where
no stationary patterns can be realized. Here all possible patterns appeared
during the system evolution are transient and at final stages the system is
totally homogeneous. According to the behavior of both $\langle x\rangle$ and
$\langle(\delta x)^2\rangle$ one can say that the average goes toward uniform
stationary value $x_0=x_{HD}$, whereas the dispersion increases at early stages
(formation of transient patterns) and after it decreases toward zeroth values
(no dispersion in adatoms concentration field is realized). Therefore, the
systems moves into homogeneous state. The situation is crucially changed when
the nonequilibrium chemical reactions are introduced. Here such reactions
freeze patterns formed at coarsening stage, and patterns develop very slowly.
As a results of competing between chemical reactions and potential interactions
between atoms such patterns are stable in time and can be considered as
stationary ones (see Fig.\ref{evol}). Indeed, here the averaged value $\langle
x\rangle$ takes lower values than $x_0$ related to $\beta=0$ and
the dispersion $\langle(\delta x)^2\rangle$ does not decrease in time at late
stages. Therefore, the quantity $\langle(\delta x)^2\rangle$ can be used an
effective order parameter. Indeed, if it decreases in time toward zero, then
the system moves into homogeneous state, whereas nonzeroth values for the
dispersion mean that modulated spatial patterns are realized. From
Fig.\ref{avX} one can see that at $\beta\ne 0$ at large adsorption rate
$\alpha$ the system is in the high density state, whereas at elevated
interaction strength of the adsorbate small islands of the dense phase are
possible; here due to large interactions between adsorbate the
evaporation/dissolution processes are less probable. Comparing curves related
to two cases $\tau=0$ and $\tau\ne0$ one can find that in the last case both
$\langle x\rangle$ and $\langle(\delta x)^2\rangle$ exhibit nonmonotonic
behavior, here oscillations with small amplitude are possible. It is well
related to results of the linear stability analysis. Moreover, it is
interesting to note that in the case $\tau=0$ most of islands of the dense
phase are of equiaxial symmetry, whereas at $\tau\ne0$ such islands are
elongated in one of two possible equivalent directions.

\begin{figure*}[!t]
 a) \includegraphics[width=0.45\textwidth]{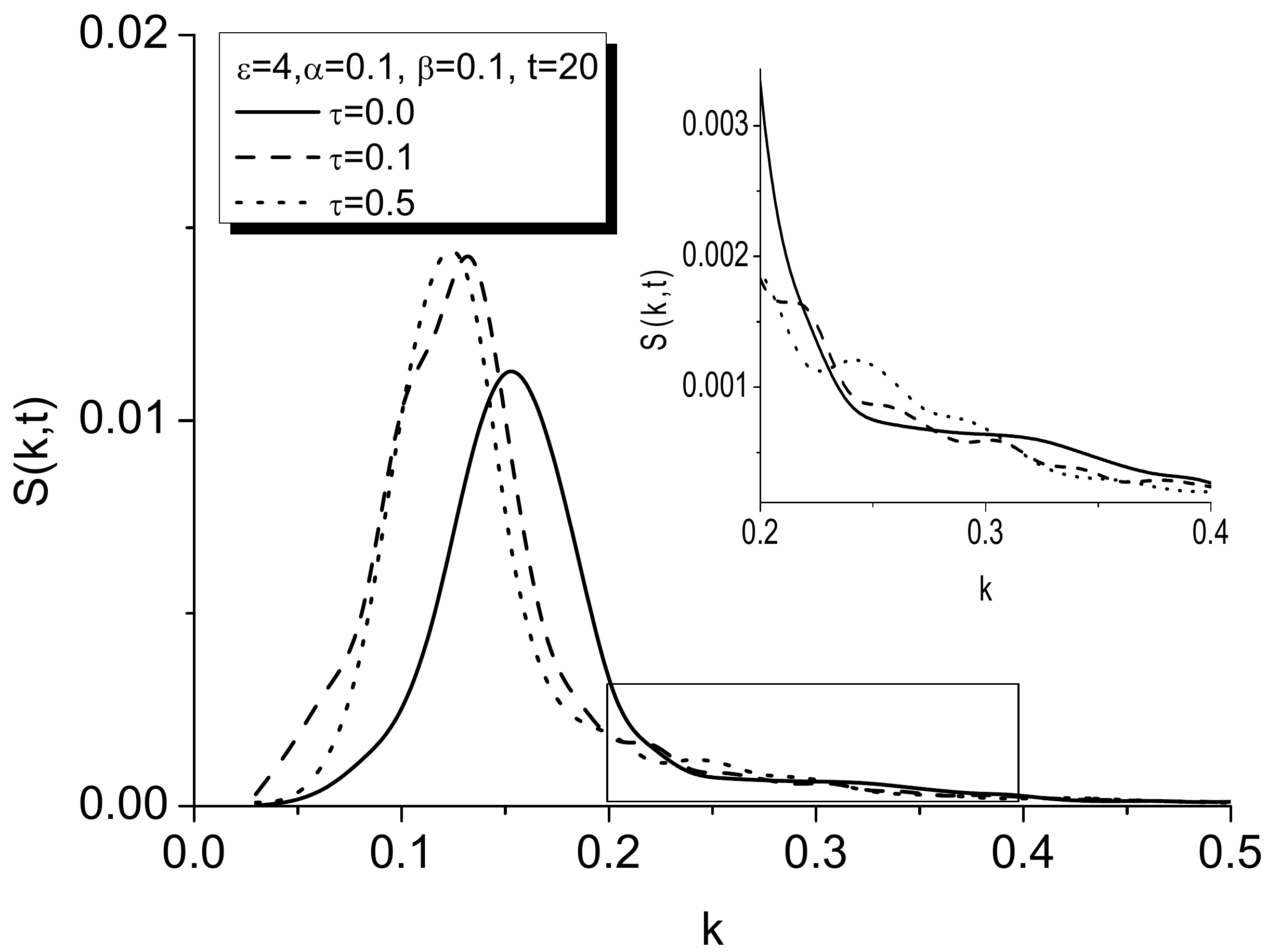}
 b) \includegraphics[width=0.45\textwidth]{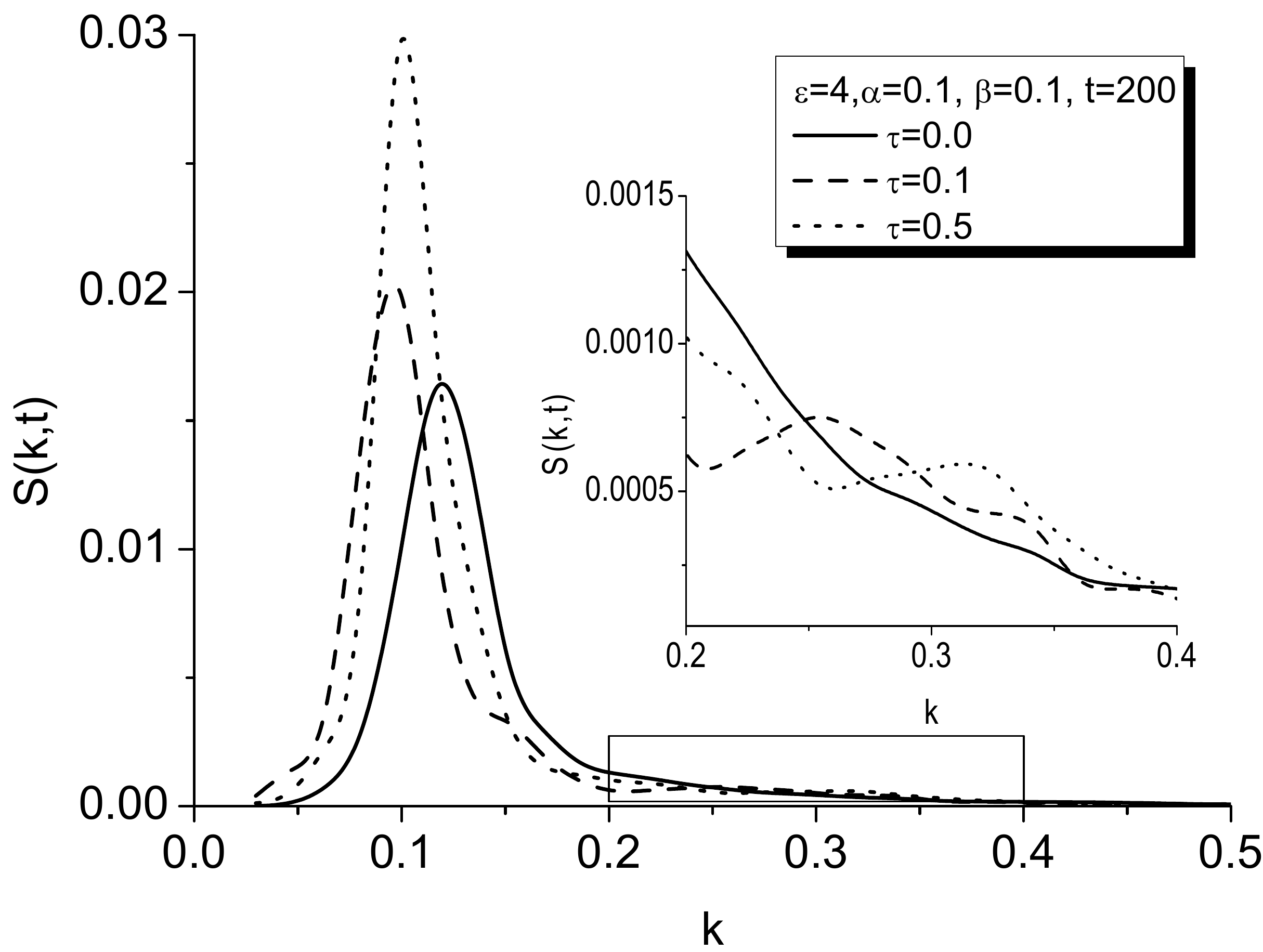}
\caption{Structure function dynamics for different values of the diffusion flux
relaxation time $\tau$ at $t=20$ (a) and $t=200$ (b). Other parameters are:
$\alpha=0.1$, $\beta=0.1$, $\varepsilon=4$\label{S(kt)}}
\end{figure*}
Let us consider dynamics of the structure function at different values for
$\tau$ and different time intervals (see Fig.\ref{S(kt)}). To calculate
$S(k,t)$ we have used fast Fourier transformation procedure. Let us start with
the simplest case of $\tau=0$ (see solid lines in Figs.\ref{S(kt)}a,b). Here
only major peak of $S(k,t)$ is realized that is related to period of islands.
There is smooth behavior of the structure function tails. During the system
evolution the peak is shifted toward stationary value of the island size; its
height increases (islands become well defined and boundaries between dense and
diluted phases become less diffusive). In the case
\begin{figure*}
\centering a)\includegraphics[width=0.45\textwidth]{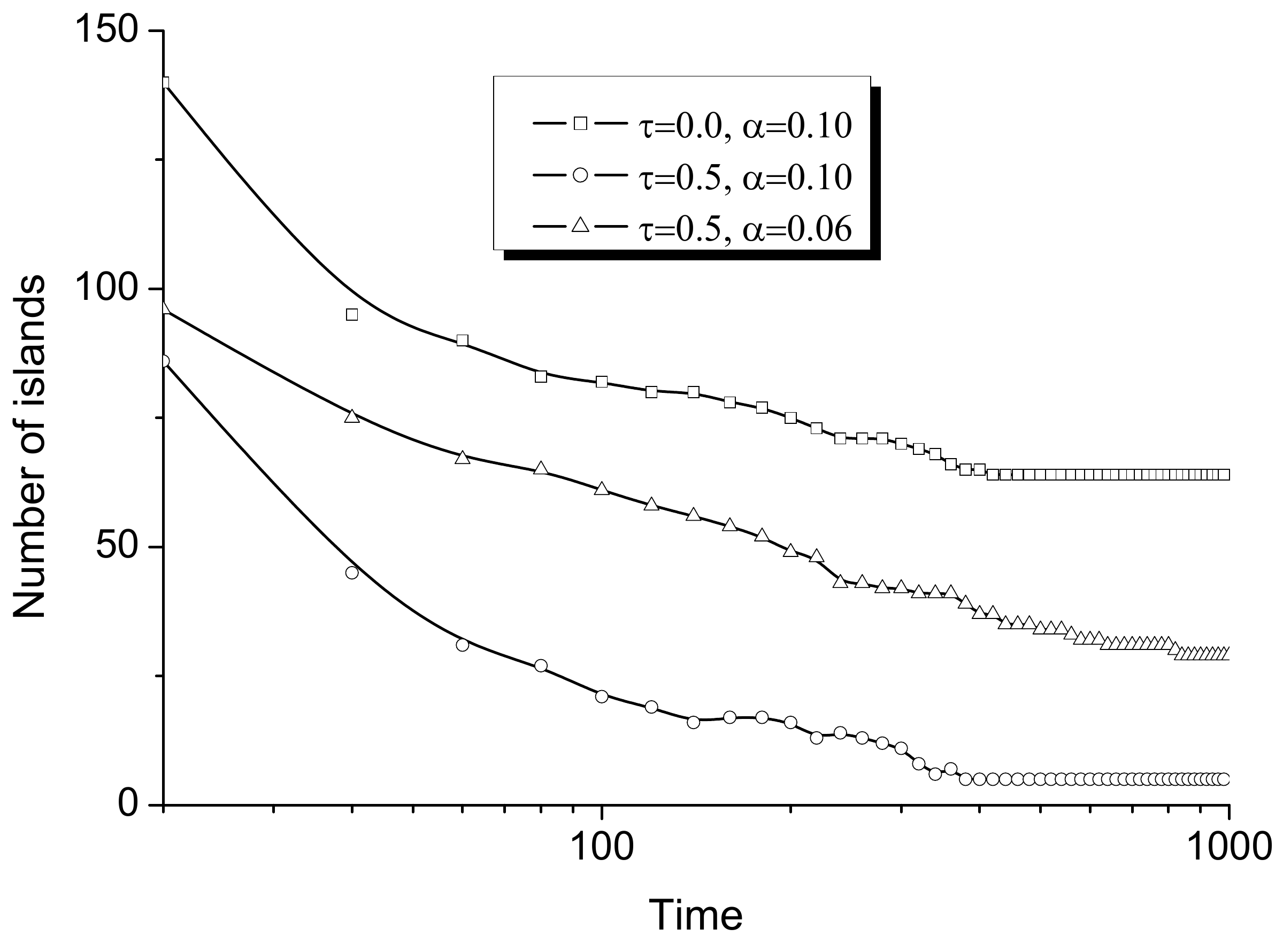}
b)\includegraphics[width=0.45\textwidth]{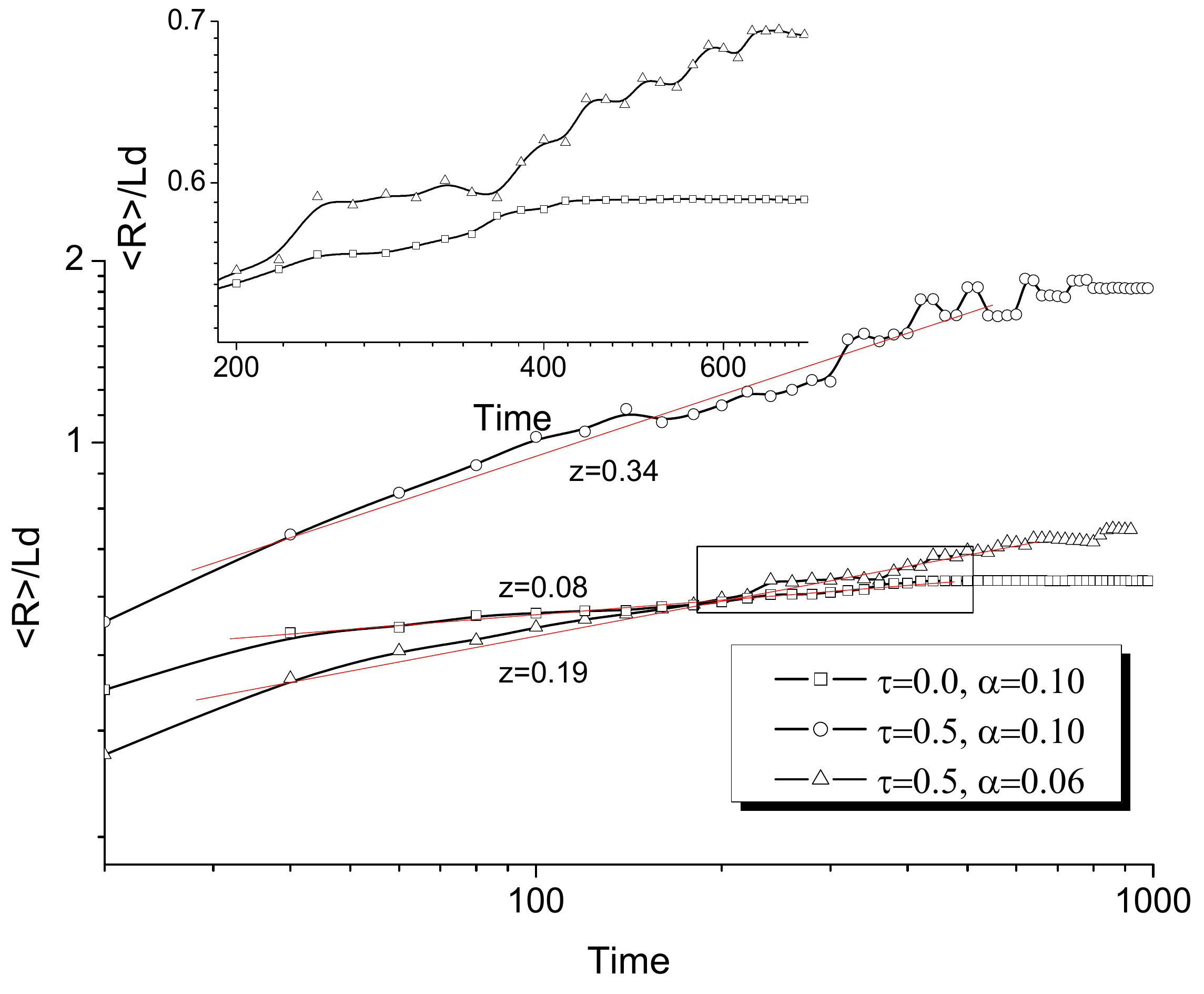}\\
c)\includegraphics[width=0.45\textwidth]{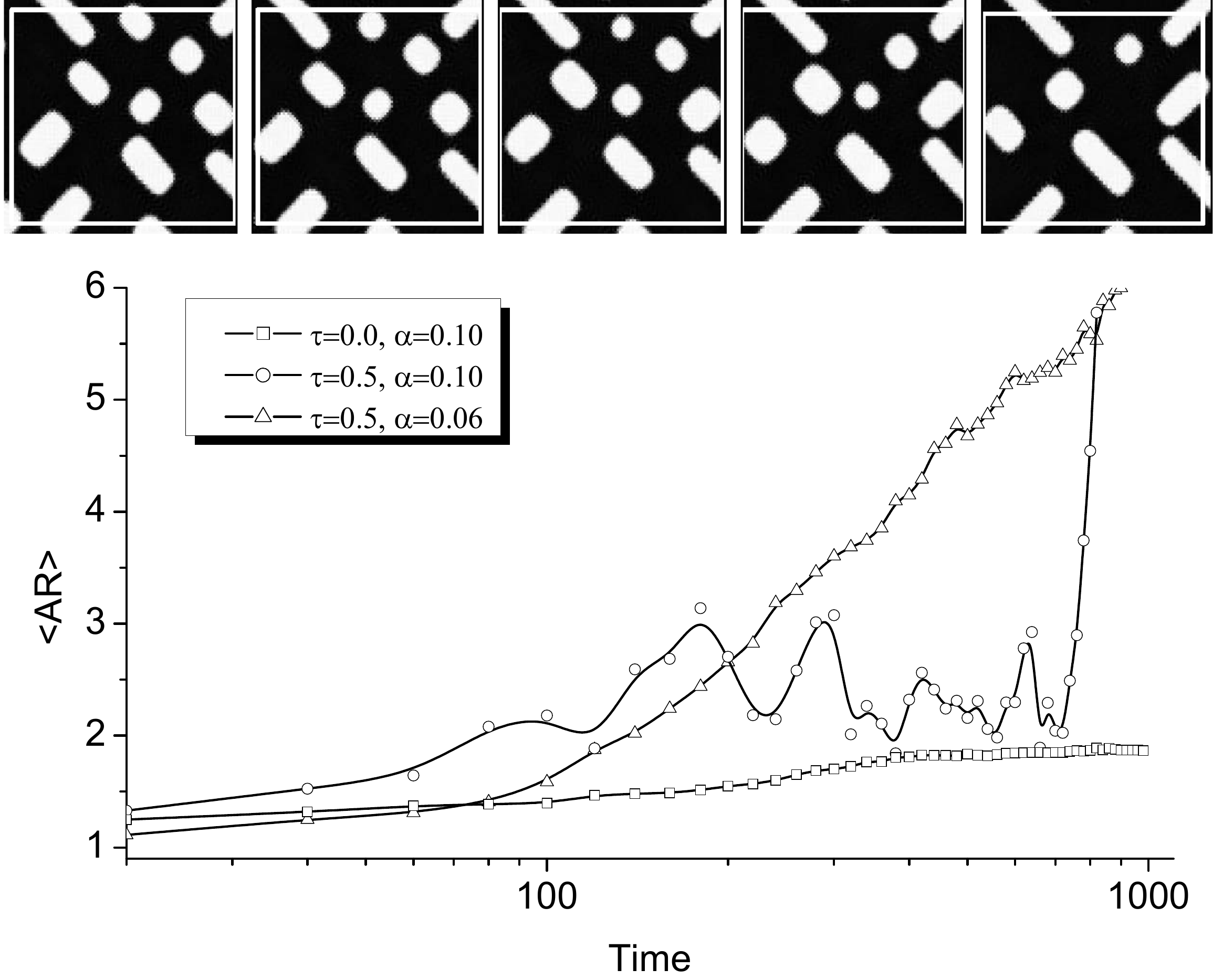}
 \caption{Dynamics of the number of islands (a),  averaged
island size (b) and the corresponding averaged aspect ratio (c) at $\beta=0.1$.
Values for the dynamical exponent $z$ related to the scaling law $\langle
R\rangle\propto t^z$ are shown near the approximate lines. In plot (c) snapshot
of pieces of the system obtained at $t=200$, 220, 240, 280, 300 illustrate a
change of $R_x/R_y$; other parameters are: $\varepsilon=4$, $\beta=0.1$,
$\alpha=0.1$.\label{R(t)}}
\end{figure*}
$\tau\ne 0$ we get one major peak at small wave number and additional peaks at
large $k$. Emergence of such minor peaks means formation of other patterns
(patterns with other periods). In the course of time an amplitude of such
satellite peaks decreases that means pattern selection processes when the
system selects one most unstable mode characterized by the major peak whose
height increases. This oscillatory behavior of the structure function at large
wave numbers is well predicted by the linear stability analysis.

Next, let us study a behavior of the averaged radius of islands. In our
computations we have calculated number of sites related to one island. This
number corresponds to a square of the island. Assuming that an island with
spherical symmetry has the same square we have computed the radius $\langle R
\rangle$ of the corresponding spherical island. In Fig.\ref{R(t)}a we plot
dynamics of the number of islands at different values for $\tau$, $\varepsilon$
and $\alpha$. Dynamics of the quantity $\langle R \rangle$ (measured in units
of diffusion length $L_d$) is shown in Fig.\ref{R(t)}b. It follows that the
system attains the stationary state with finite number of islands. Comparing
curves related to different $\tau$, one can find that nonequilibrium effects
related to $\tau\ne 0$ decrease essentially the number of islands (but
stationary islands have large size). If the adsoprtion rate $\alpha$ is small,
then number of islands is large (they are characterized by small sizes). From
Fig.\ref{R(t)}b it is seen that in the simplest case of pure dissipative system
characterized by $\tau=0$ the averaged radius is a monotonically increasing
quantity. At large time interval it attains a stationary value and does not
changed. It means that the coarsening procedure is finished and we get
stationary patterns. It follows that islands of the adsorbate have the averaged
length $\langle R\rangle\sim (0.15\div 0.9)L_d$. Following
Refs.\cite{CTT06,CTT07} one can estimate $\langle R \rangle$ considering
deposition of Al on TiN(100): at room temperature one has the lattice constant
$a_{Al}=4.05\times10^{-10}m$, the pair interaction energy $\epsilon_0=-0.22~eV$
with the coordination number $Z=4$ gives $r_0=Z a_{Al}\simeq 1.6\times
10^{-9}m$, the diffusion constant $D=10^{-10}cm^2s^{-1}$. Hence the patterns
have the size $\langle R\rangle\sim (10\div 60)\times 10^{-9}m$.

\begin{figure}[!ht]
\centering
 a)
 \includegraphics[width=0.45\textwidth]{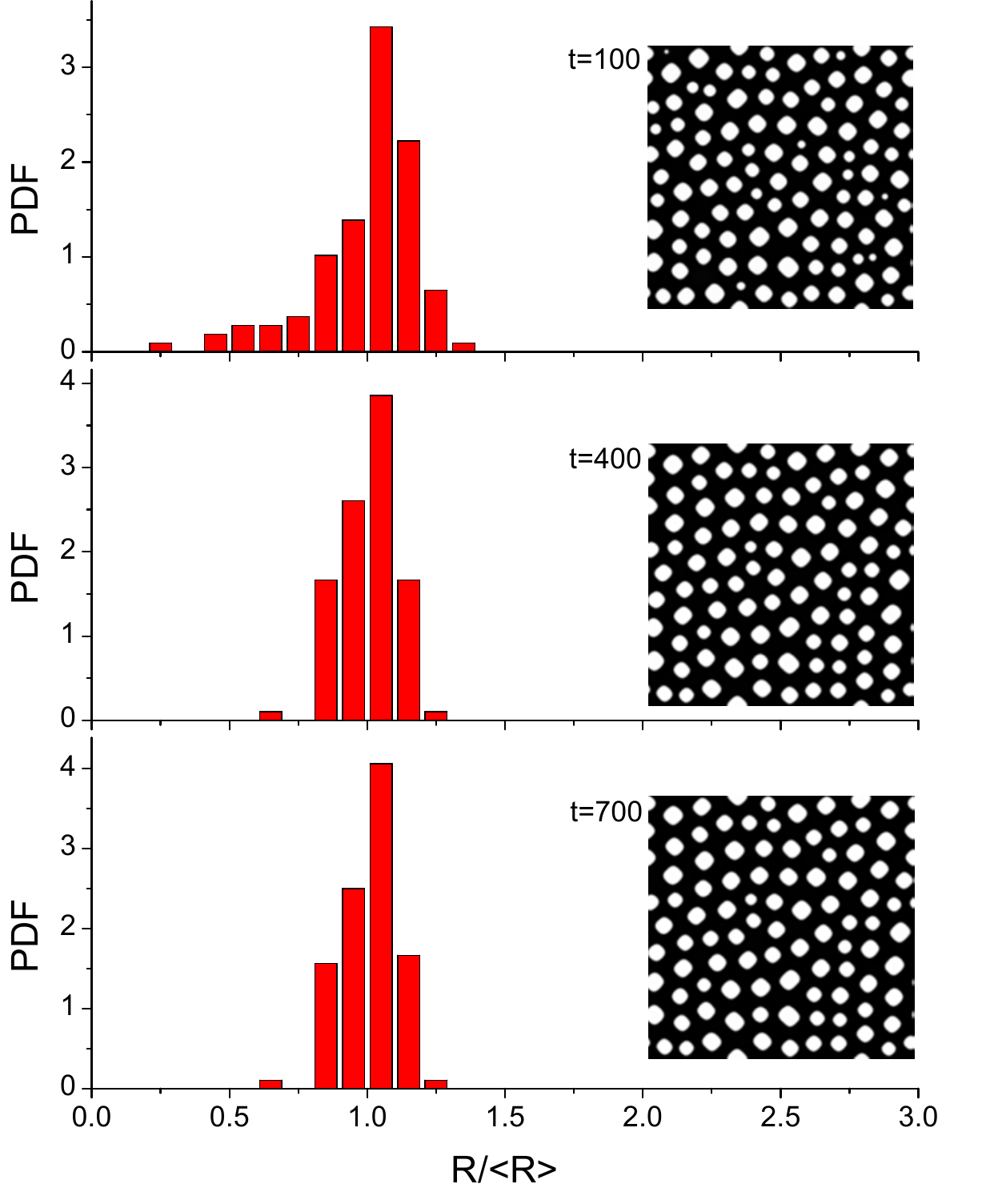}
 b) \includegraphics[width=0.45\textwidth]{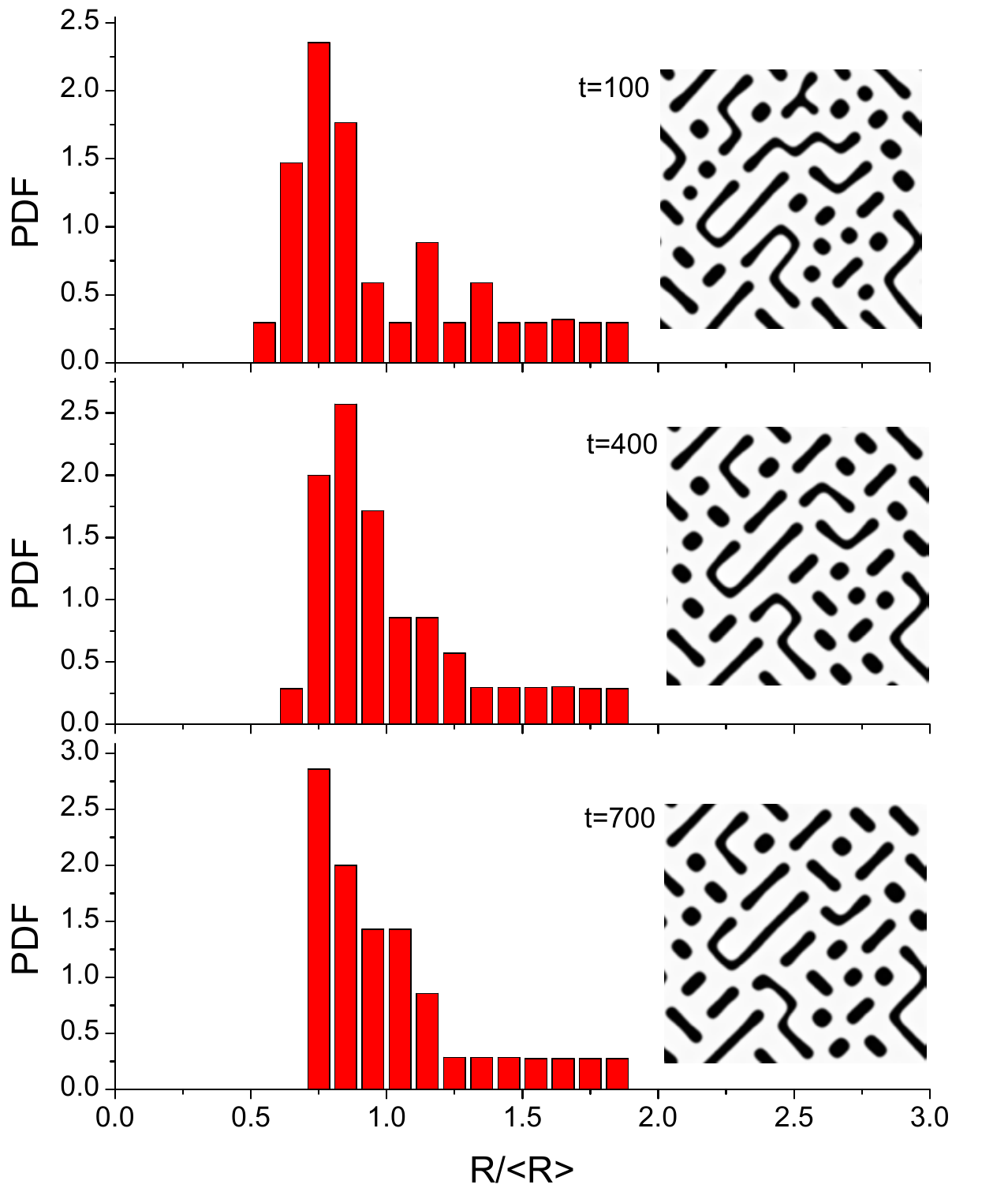}\\
c) \includegraphics[width=0.45\textwidth]{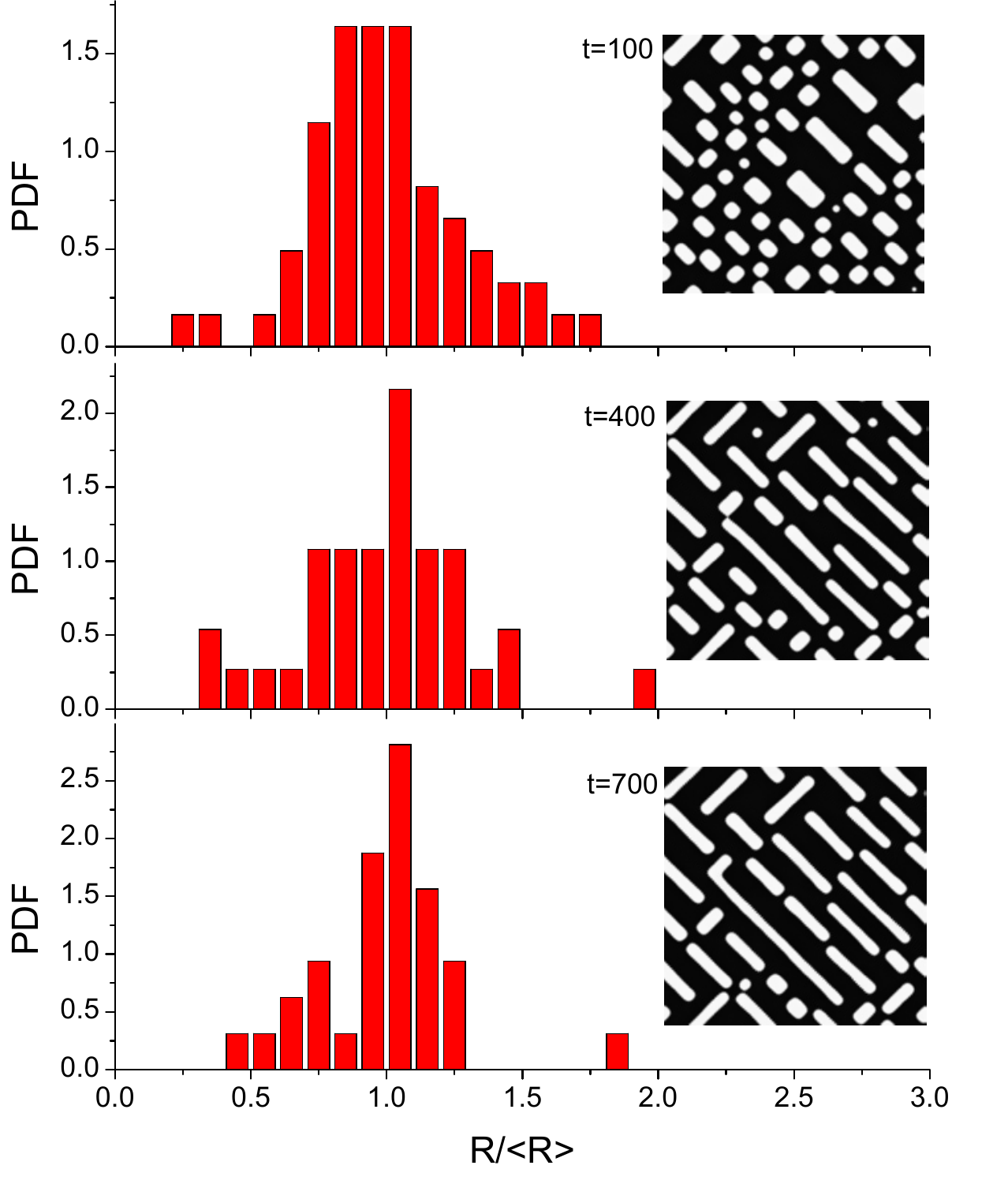}
 d)\includegraphics[width=0.45\textwidth]{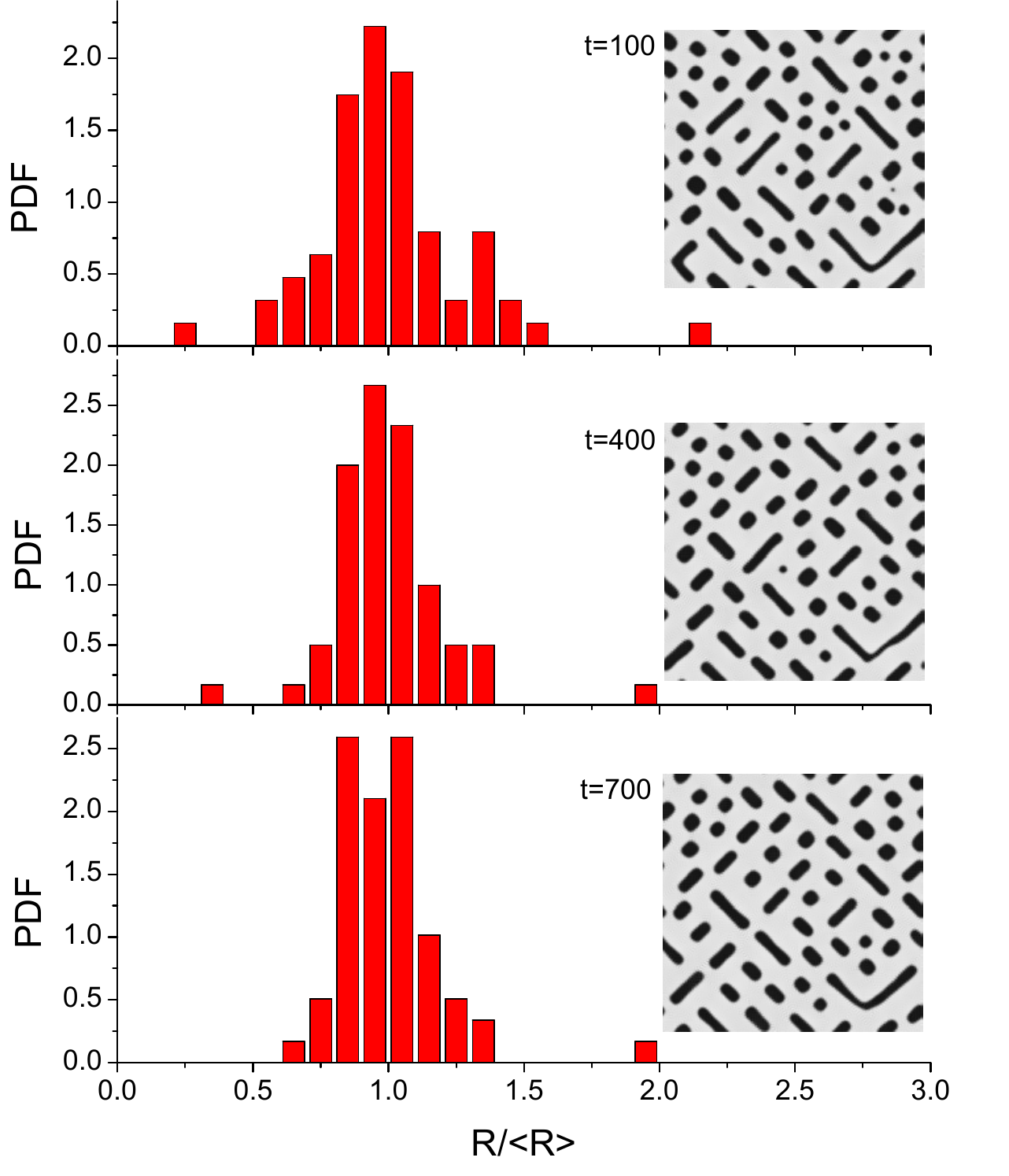}
 \caption{Evolution of the probability density function of the
dimensionless radius of the islands size at different $\tau$ at $\beta=0.1$: a)
$\tau=0$, $\varepsilon=4$, $\alpha=0.06$; b) $\tau=0$, $\varepsilon=3$,
$\alpha=0.25$; c) $\tau=0.5$, $\varepsilon=4$, $\alpha=0.06$; d) $\tau=0.5$,
$\varepsilon=3$, $\alpha=0.25$\label{PDF}}
\end{figure}

Studying scaling properties of the island growth we have estimated a scaling
exponent $z$ in the islands size growth law $\langle R\rangle\propto t^z$.
Fitting the data related to stage of growth, we have obtained $z=0.08$ for pure
dissipative system with $\varepsilon=4$. Taking $\tau\ne0$, one can find that
at small $\varepsilon$ (in our case $\varepsilon=4$) the averaged radius of
islands increases in nonmonotonic manner. Fitting the scaling regime we have
found that at $\tau=0.5$ the scaling exponent takes large values, in our case
$z=0.34$. When the adsorption rate decreases the growth processes delay and
characterize by small dynamical exponent $z$ (at $\alpha=0.06$ one has
$z=0.19$). Therefore, nonequilibrium effects related to relaxation of the
diffusion flux characterized by $\tau\ne0$ accelerate growth processes, the
same is observed when adsorption rate increases. It is interesting to note that
at $\tau\ne0$ oscillations predicted for the averaged field and the structure
function are possible for the dependence $\langle R(t)\rangle$. It means that
islands can change their size during the system evolution in oscillating
manner. Amplitude of oscillations is well pronounced for large $\alpha$ and
small $\varepsilon$.

To prove that islands can oscillatory change their sizes we compute an averaged
aspect ratio $\langle AR\rangle=\langle R_x/R_y\rangle$, where $R_x$ and $R_y$
are sizes of an island in $x$- and $y$-direction, respectively. From the
obtained dependencies shown in Fig.\ref{R(t)}c it follows that for pure
dissipative system deviations from the straight line are small and may be
considered as fluctuations in $R_x$ and $R_x$ values. At $\tau\ne0$ and small
$\varepsilon$ oscillations in $\langle AR\rangle$ are well pronounced. Such
oscillations in lateral and longitudinal sizes of islands mean that at some
fixed time interval most of the islands grow in one direction whereas in other
direction their size decreases, in next time interval these two directions are
changed. When this scenario is repeated one gets an oscillation picture of
island size growth. In Fig.\ref{R(t)}c we present snapshots showing change of
the growth orientation during the system evolution.

In our study we are interesting also in a behavior of the probability density
functions (PDF) of the island sizes distribution. Let us consider PDF's for
pure dissipative system, initially. As Figs.\ref{PDF}a,b shows islands of
different sizes formed at initial stages evolves in such manner that the most
probable value for $R$ is located around $\langle R\rangle$. Islands of other
sizes are possible with small probabilities. In the stationary limit PDF has
unique well pronounced peak, small fluctuations of island size distribution are
possible and are shown as satellite peaks around major one. In the case of
$\tau\ne 0$ (see Figs.\ref{PDF}c,d) the hyperbolic transport promotes formation
of islands with different sizes. The principle difference comparing to the case
of pure dissipative system is in a bimodal form of PDF observed at large times.
Here at early stages islands of different sizes are formed, and during the
system evolution following the Ostwald ripening mechanism small islands
dissolve and large islands reduces their sizes. In such a case two well
pronounced major peaks of PDF indicate that there two most probable sizes of
islands related to $R>\langle R\rangle$ and $R<\langle R\rangle$.

\section{Conclusions}
We have studied dynamics of islands formation of the adsorbate using
generalized approach including persistent motion of particles having finite
speed at initial stages and diffusion kinetics at final ones. It was found that
stabilization of nano-patterns in such class of reaction-Cattaneo models is
achieved by noneqilibrium chemical reactions. It was found that during the
system evolution pattern selection processes are realized. We have shown that
possible oscillatory regimes for islands formation are realized at finite
propagation speed related to nonzero relaxation time for the diffusion flux.

Our results can be used to describe formation of nano-islands at processes of
condensation from the gaseous phase. Despite we have considered a general model
where relaxation time $\tau_J$ for the diffusion flux is small but nonzeroth
value, one can say that condensation processes with formation of metallic
islands can be described in the limit $\tau_J/\omega_D^{-1}\lesssim 10^{-3}$
(here $\omega_D$ is the Debye frequency), whereas nano-islands formation with
$\tau_J/\omega_D^{-1}\sim 10^{-1}\div 10^{-2}$ is possible for soft matter
condensation (semiconductors, polymers, etc.).



\begin{thebibliography}{00}

\bibitem{ZTWE96} T.Zambelli, J.Trost, J.Wintterlin, G.Ertl, Phys.rev.Lett. 79 (1996) 795.

\bibitem{GLRBE94}V.Gorodetskii, J.Lauterbach, H.A.Rotermund, J.H.Block, G.Ertl, Nature 370 (1994) 276.

\bibitem{KNSZGC91} K.Kern, H.Niehus, A.Schatz, P.Zeppenfeld, J.George, G.Cosma,
Phys.Rev.Lett. 67 (1991) 855.

\bibitem{PWCL97}T.M.Parker, L.K.Wilson, N.G.Condon, F.M.Leissle, Phys.Rev.B 56 (1997)
6458.

\bibitem{BGBK98}H.Brune, M.Giovannini, K.Bromann, K.Kern, Nature 394 (1998)
451.

\bibitem{CF99}P.G.Clark, C.M.Friend, J.Chem.Phys. 111 (1999) 6991.

\bibitem{Nature99}K.Pohl, M.C.Bartelt, J.de la Figuera, N.C.Bartelt, J.Hrbek, R.Q.Hwang, Nature 397 (1999)
238.

\bibitem{ZKRGL94} P.Zepenfeld, M.Krzyzowski, C.Romainczyk, G.Gomsa,
M.G.Lagally, Phys.Rev.Lett. 72 (1994) 2737.

\bibitem{M91}V.I.Marchenko, JETP Lett. 67 (1991) 855.

\bibitem{V92}D.Vanderbit, Surf.Sci. 268 (1992) L300.

\bibitem{202} Y.W.Mo, B.S.Swartzentruber, B.Kariotis, M.B.Webb, M.G.Lagally, Phys.Rev.Lett. 63 (1989)
2393.

\bibitem{sem2002} G.E.Cirlin, V.A.Egorov, L.V.Sokolov, P.Werner,
Semiconductors 36 (2002) 1294-1298.

\bibitem{206} J.P.Bucher, E.Hahn, P.Fernandez, C.Massobrio, K.Kern, Europhys.Lett. 27 (1994)
473.

\bibitem{MBE1998} Harald Brune, Surf.Sci.Reports 31 (1998) 121-229.

\bibitem{Binder} K.Binder, in: P.Haasen (Eds.), Material Science and Technology: Phase Transformations in
Materials, Vol.5, VCH, Weinham, 1990.

\bibitem{TH96} Q.Tran-Cong, A.Harada, Phys.Rev.Lett. 76 (1996) 1162.

\bibitem{HM96}M.Hildebrand, A.S.Mikhailov, J.Phys.Chem. 100 (1996) 19089.

\bibitem{BHKM97}D.Batogkh, M.Hildebrant, F.Krischer, A.Mikhailov, Phys.Rep. 288 (1997) 435.

\bibitem{HME98_1}M.Heldebrand, A.S.Mikhailov, G.Ertl, Phys.Rev.Lett. 81 (1998) 2602(4).

\bibitem{HME98_2} M.Heldebrand, A.S.Mikhailov, G.Ertl, Phys.Rev.E 58 (1998) 5483(11).

\bibitem{ME94}A.Mikhailov, G.Ertl, Chem.Phys.Lett. 238 (1994) 104.

\bibitem{MW2005} Sergio E.Mangioni, Horacio S. Wio, Phys.Rev.E. 71 (2005) 056203.

\bibitem{M2010} Sergio E.Mangioni, Physica A 389 (2010) 1799.

\bibitem{PhysD2009} D.O.Kharchenko, S.V.Kokhan, A.V.Dvornichenko, Physica D 238 (2009)
2251.

\bibitem{PhysScr2011} Dmitrii O.Kharchenko, Vasyl O.Kharchenko, Irina
O.Lysenko, Phys.Scr. 83 (2011) 045802(9)

\bibitem{JP89}D.D.Joseph, L.Preziosi, Rev.Mod.Phys. 61 (1989) 41.

\bibitem{H1999} Werner Horsthemke, Phys.Rev.E 60 (1999) 2651.

\bibitem{lzg} N.~Lecoq, H.~Zapolsky, P.~Galenko, Eur.Phys.Jour.ST. 177 (2009) 165.

\bibitem{PhysA1_2010} P.K.Galenko, Dmitrii Kharchenko, Irina Lysenko, Physica
A 389 (2010) 3443.

\bibitem{CEJP} Dmytro Kharchenko, Vasyl Kharchenko, Irina Lysenko,
Cent.Eur.J.Phys. 9(3) (2011) 698.

\bibitem{Ghosh2010} Pushpita Ghosh, Shrabani Sen, Deb Shankar Ray, Phys.Rev.E 81 (2010) 026205.

\bibitem{CWM2002} S.B.Casal, H.S.Wio, S.Mangioni, Physica A 311 (2002) 443.

\bibitem{PRE822010}Sergio E.Mangioni, Roberto R.Deza, Phys.Rev.E 82 (2010) 042101.

\bibitem{CTT06} M.G.Clerc, E.Tirapegui, M.Trejo, Phys.Rev.Lett. 97 (2006) 176102.

\bibitem{CTT07} M.G.Clerc, E.Tirapegui, M.Trejo, Eur.Phys.J. ST. 146 (2007) 407.


\bibitem{Grant200220061} K.R.Elder, Mark Katakowski, Mikko Haataja, Martin Grant,
Phys.Rev.Lett. 88 (2002) 245701.

\bibitem{Grant200220062} K.R.Elder, Martin Grant, Phys.Rev.E 70 (2004) 051605.

\bibitem{Grant200220063} Peter Stafanovich, Mikko Haataja, Nikolas Provatas, Phys.Rev.Lett. 96 (2006)
225504.

\bibitem{PhysA2010} D.Kharchenko, I.Lysenko, V.Kharchenko, Physica A 389 (2010) 3356.


 \end{thebibliography}
\end{document}